\documentclass[12pt,preprint]{aastex}
\begin{document}

\newcommand{\lya}{Lyman-$\alpha$}
\newcommand{\eqw}{\hbox{EW}}
\def\erg{\hbox{erg}}
\def\cm{\hbox{cm}}
\def\sec{\hbox{s}}
\def\f17{f_{17}}
\def\Mpc{\hbox{Mpc}}
\def\cMpc{\hbox{cMpc}}
\def\Gpc{\hbox{Gpc}}
\def\nm{\hbox{nm}}
\def\km{\hbox{km}}
\def\kms{\hbox{km s$^{-1}$}}
\def\year{\hbox{yr}}
\def\Myr{\hbox{Myr}}
\def\Gyr{\hbox{Gyr}}
\def\deg{\hbox{deg}}
\def\arcsec{\hbox{arcsec}}
\def\microJy{\mu\hbox{Jy}}
\def\zre{z_r}
\def\fesc{f_{\rm esc}}
\def\lstar{\ifmmode {L_\star}\else
                ${L_\star}$\fi}
\def\phistar{\ifmmode {\phi_\star}\else
                ${\phi_\star}$\fi}

\def\ergcm2s{\ifmmode {\rm\,erg\,cm^{-2}\,s^{-1}}\else
                ${\rm\,ergs\,cm^{-2}\,s^{-1}}$\fi}
\def\ergsec{\ifmmode {\rm\,erg\,s^{-1}}\else
                ${\rm\,ergs\,s^{-1}}$\fi}
\def\kmsMpc{\ifmmode {\rm\,km\,s^{-1}\,Mpc^{-1}}\else
                ${\rm\,km\,s^{-1}\,Mpc^{-1}}$\fi}
\def\kpc{{\rm kpc}}
\def\taulya{\tau_{Ly\alpha}}
\def\taubar{\bar{\tau}_{Ly\alpha}}
\def\llya{L_{Ly\alpha}}
\def\ldlya{{\cal L}_{Ly\alpha}}
\def\nbar{\bar{n}}
\def\Msun{M_\odot}
\def\sqamin{\Box'}


\title{Luminosity functions of Lyman-$\alpha$ emitters at Redshift z=6.5 
and z=5.7: Evidence against reionization at $z \approx 6$}

\author{
Sangeeta Malhotra and James E. Rhoads 
}
\affil{Space Telescope Science Institute, 3700 San Martin Dr, Baltimore, MD21218}
\email{san@stsci.edu, rhoads@stsci.edu}

\begin{abstract}


\lya\ emission from galaxies should be suppressed completely or
partially at redshifts beyond reionization. Without knowing the
intrinsic properties of galaxies at $z = 6.5$ this attenuation is
hard to infer in any one source, but can be infered from a comparison
of luminosity functions of \lya\ emitters at redshifts just before and
after reionization. We combine published surveys of widely varying
depths and areas to construct luminosity functions at z=6.5 and 5.7,
where the characteristic luminosity \lstar\ and density \phistar\ are
well constrained while the faint-end slope of the luminosity function
is essentially unconstrained. Excellent consistency is seen in all but
one published result.  We then calculate the likelihood of obtaining
the z=6.5 observations given the z=5.7 luminosity function with (A) no
evolution and (B) an attenuation of a factor of three. Hypothesis (A)
gives an acceptable likelihood while (B) does not.  This indicates
that the $z=6.5$ \lya\ lines are not strongly suppressed by a neutral
intergalactic medium and hence that reionization was largely complete
at $z\approx 6.5$.

\end{abstract}



\section{Introduction}

The epoch of reionization marks a phase transition in the universe,
when the intergalactic medium was ionized. Recent observations of
$z>6$ quasars show a Gunn-Peterson trough, implying that the
reionization of intergalactic hydrogen was not complete until
$z\approx 6$ (Becker et al 2001, Fan et al 2002).  Yet microwave
background observations imply substantial ionization as early as $z\ga
15$ (Spergel et al 2003).  Either reionization took place at $z\approx
6$, or $z\approx 15$, or occurred twice (e.g., Cen 2003).  \lya\
emitting galaxies offer another, independent test of reionization,
because they are sensitive to neutral fraction of about $\sim 0.1$,
rather than $\sim 0.01$ for Gunn-Peterson effect.

Because \lya\ photons are resonantly scattered by neutral hydrogen, we
expect \lya\ line fluxes to be strongly attenuated for sources at rest
in an intergalactic medium (IGM) that has a substantial neutral
fraction (Miralda-Escud\'{e} 1998, Loeb \& Rybicki 1999,
Haiman \& Spaans 1999).  To zeroth
order this should produce a decrease in \lya\ galaxy counts at
redshifts beyond the end of hydrogen reionization. A fraction of the line
flux may escape due to velocity structure in the line and/or local
ionization of the IGM by the galaxy producing the line, but an
effective optical depth of at least $1$ to $2$ is unavoidable in any
model with a neutral IGM (Haiman 2002; Santos 2004).

By now a fair number of \lya\ sources have been observed at $z>6$ (Hu
et al 2002; Kodaira et al 2003; Rhoads et al 2004, Kurk et al. 2004,
Stern et al. 2004) and even up to z=10 (Pello et al. 2004). In each
individual case we cannot say whether the \lya\ flux has been
attenuated by a factor of $\ga 3$ by the IGM, since we do not know the
intrinsic \lya\ flux.  Thus, while the evidence to date supports a
largely ionized IGM at $z\approx 6.5$ (Rhoads et al 2004), we cannot
completely exclude a neutral IGM if these sources had been
intrinsically brighter than the observations would have us
believe.

This difficulty can be overcome if we compare sizeable samples of
galaxies before and after reionization.  We previously developed such
an approach to demonstrate that the IGM at $z\approx 5.7$ is largely
ionized (RM01), despite the large increase in IGM
opacity with redshift seen in quasar spectra at $z\approx 5.7$
(Djorgovski et al 2001).  Subsequent work has shown that the
true Gunn-Peterson trough likely sets in at $z\ga 6.2$ (Becker et
al 2001, Pentericci et al 2002, Fan et al 2002).

In this {\it Letter},  we assemble presently available data 
at $z\approx 6.5$, before the end of the reionization era
suggested by the GP trough.  We compare these data with a control
sample at  $z\approx 5.7$.  The $z\approx 5.7$ window is
the closest atmospheric window to $z\approx 6.5$ and hence affords
high sensitivity and a short time baseline (thus minimizing possible
effects due to galaxy evolution) while still spanning the end of the
reionization era inferred from the Gunn-Peterson test. We construct a
luminosity function at redshift 
z=5.7 (\S~\ref{lf5sec})  and at redshift 6.5 (\S~\ref{lf6sec}). 
We then test whether the z=6.5 \lya\
emitters are consistent with one of the following two scenarios: (A)
no change in luminosity function between z=5.7 and 6.5, or (B) a
factor of 3 reduction in the characteristic luminosity \lstar going
from z=5.7 to z=6.5 (\S~\ref{rei_sec}). In \S~\ref{sfr_sec}, we
derive the star-formation rate at z=6.5 and how that constrains the
luminosity function.

\section{The $z=5.7$ \lya\ Luminosity Function} 
\label{lf5sec}

We base our $z=5.7$ luminosity function (LF) fit on a combination of four
surveys: The Large Area Lyman Alpha (LALA) survey (Rhoads \& Malhotra
2001, Rhoads et al 2003); Hu et al (2004); Ajiki et al (2004); and
Santos et al (2004).  Properties of these surveys are summarized in
table~\ref{lf5tab}. The resulting LF is also consistent with the upper
limits presented by Martin \& Sawicki (2004).

In the LALA survey, we have previously published a sample of 18
narrowband-selected $z\approx 5.7$ \lya\ candidates (Rhoads \&
Malhotra 2001).  Early followup observations confirmed three of the
first four that were spectroscopically observed (Rhoads et al 2003),
suggesting a total sample of $\sim 13$ after correction for
reliability. 
The uncertainty assigned to the \lya\ sample size when fitting LFs
must account for uncertainties both in the candidate counts and
the spectroscopically determined reliability.  Given a sample of
four spectroscopic targets and three confirmations, 
the variance in the number confirmed follows that of a binomial
distribution, $N p (1-p) = 4\times (3/4) \times (1/4) = 3/4$
for $N=4$ trials and  probability $p=3/4$ for confrimation.
This gives a fractional uncertainty of $29\%$.  Combining this
with the $\sqrt{1/18}$ fractional uncertainty in the candidate counts
gives a final fractional uncertainty of $0.37$.  The Poisson sample
size giving the same fractional uncertainty is $N_{equiv} = 7.3$.
Thus, when fitting the luminosity function using Cash statistics
(see below), we treat this sample as having discovered $7.3$ sources
in a volume , and
in our luminosity function fits we treat this survey as having
detected 7.3 sources in an effective volume $54\% = 7.3/(18*3/4)$ of
the true survey volume.

\begin{figure} \epsscale{0.8} 
\plotone{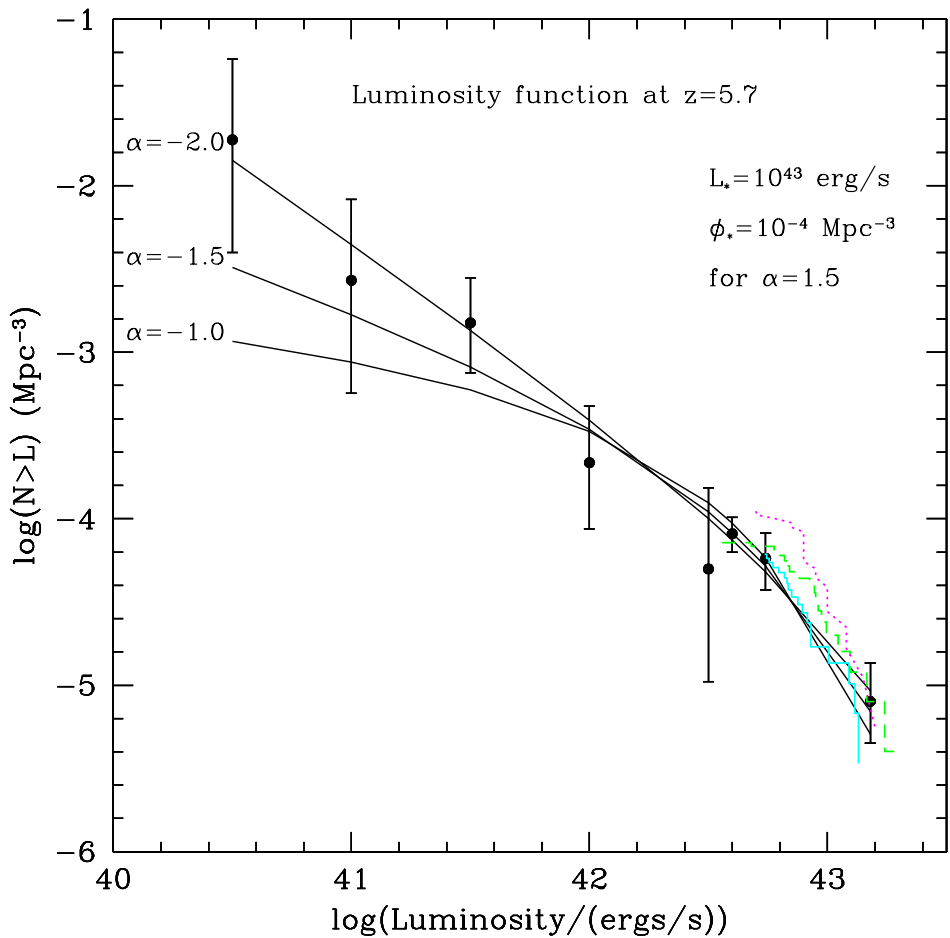} 
\caption{\lya\
luminosity function at redshift $z\approx 5.7$.  Data points with
error bars show the measured densities above detection threshold and
their corresponding uncertainties for samples from Table~\ref{lf5tab}.  Three
curves are shown, corresponding to the best Schechter function fits to
the data for assumed faint-end slopes $\alpha = 1$, 1.5, and~2 (from
bottom to top). Cumulative flux-number density curves are shown for
the LALA survey (RM01 and Rhoads et al. 2003) - solid line , 
Hu et al. 2004 (dashed line), and Ajiki et al. 2003 - dotted line, and show consistency with each other.}
\label{z57lumfn}
\end{figure}

The Hu et al (2004) sample consists of $19$ spectroscopically
confirmed $z\approx 5.7$ \lya\ galaxies. They obtained this sample
from 23 spectra among a list of 26 candidates selected 
from narrowband imaging.  We calculate their line luminosities using published
narrow-band fluxes. Folding in the spectroscopic confirmation uncertainties
with the above formalism, we obtain a final fractional uncertainty
of 23\% in the source density, for an effective sample size 
$N_{equiv}=19.0$ and volume $88\% = 19.0/(26*19/23)$ of the
true survey volume.
Another comparable narrowband sample at $z\approx 5.7$ has
been published by Ajiki et al (2003) and is plotted in figure~\ref{z57lumfn}.
However, this is likely to be biased as it is in the field of 
a previously known $z\approx 5.74$ quasar, and we omit it from
our calculations.
More recently, Ajiki et al (2004) used an intermediate band filter to search
a larger volume in the same field to shallower depth.  They found four
$z\approx 5.7$ galaxies, all near the bright end of the
\lya\ sample luminosity range.  We use this survey in our fit, because
it provides a valuable bright-end constraint on the LF and because the number 
density bias due to the known quasar is diluted in such a large volume field.
The study by Santos et al (2004) provides the best leverage
on the faint end of the luminosity function, because 
they have obtained slit spectra of the most strongly
magnified regions near gravitational lens galaxy clusters.
This yields a very sensitive \lya\ search over a small volume 
spanning a wide range of redshift ($2.8 \la z \la 6.7$).  In
the range $4.5 \la z \le 5.6$ they have detections, which constrain
the faint end of the $z\approx 5.7$ luminosity function.  This may
introduce a slight systematic bias in the faint end of the luminosity
function slope if redshift evolution is strong. But comparison of z=4.5
and z=5.7 LALA samples shows at most weak evolution (RM01).

Using these four surveys we can constrain the luminosity function 
of the \lya\ emitters at $z=5.7 \pm 0.1$. Assuming that the errors
in the number of \lya\ emitters found are Poissonian rather than
Gaussian,  the analog to chi-square statistic is the  Cash C stastic
(Cash 1979, Holder, Haiman \& Mohr 2001):
\begin{equation}
 C = -2 ln L = -2 \Sigma_{a=1} ^{N} n_a \ln e_a -e_a -\ln n_a!
\end{equation}
Here $n_a$ and $e_a$ are respectively the observed and expected
number of counts in sample $a$ and N is the total number of samples.

If we look at figure~\ref{z57lumfn}, it is clear that the faint end
slope of the LF is constrained only by a handful of points 
from Santos et al (2004).
Therefore we have assumed a faint-end slope $\alpha$.
Taking $\alpha=1,1.5,2$ we can fit a Schechter function 
\begin{equation}
 \phi(L) dL = \phi_\star (\frac{L}{L_\star})^\alpha 
    e^{-L/L_{\star}} \frac{dL}{L_\star} ~~.
\end{equation}
For each assumed value of $\alpha$, we have minimized $C$ using a
simple grid search.  This leads to reasonably constrained 
measurements of the parameters $L_{\star}$ and
$\phi_\star$ (albeit with correlated uncertainties).
The best fit parameters for each value of $\alpha$ are given
in table~\ref{lffittab}.


\begin{deluxetable}{lllll}
\tablecolumns{5}
\tablewidth{0pc}
\tablecaption{Surveys for $z\approx 5.7$ \lya\ emitters}
\tablehead{
\colhead{Survey ID} & \colhead{Sensitivity}  & \colhead{Volume} & \colhead{Number} &\colhead{$\sigma$} \\
                    & ($\ergcm2s$)           &  (Mpc$^{-3}$)  &   &  }
\startdata
LALA (Rhoads \& Malhotra 2001)	& $1.5\times 10^{-17}$  & $2.3 \times 10^5$ &  13.5 & 5\\
Ajiki et al 2004                & $4\times 10^{-17}$  & $ 5 \times 10^5$  &  4    & 2 \\
Hu et al. 2004                  & $1.1\times 10^{-17}$  & $2.3 \times 10^5$ &  19   & 4.9 \\
Santos et al. 2004              & $1\times 10^{-17}$ &  $2 \times 10^4 $ & 1 & 1\\
	      			& $3\times 10^{-17} $ &  9200 & 1 & 1\\
	      			& $1\times 10^{-18} $ &   2000  & 3 & 1.73\\
	      			& $3\times 10^{-18} $ &   370 & 2 & 1.4\\
	      			& $1\times 10^{-19} $ &   53 & 1 & 1\\
\enddata
\tablecomments{Surveys used for determining the $z\approx 5.7$
\lya\ luminosity function.
\label{lf5tab}}
\end{deluxetable}

\section{The $z=6.5$ \lya\ Luminosity Function}
\label{lf6sec}

While the current sample of $z\approx 6.5$ \lya\ emitters is small
($\sim 6$), they were identified by searches that spanned a wide range
of survey volume (a factor of $\sim 10^4$) and sensitivity (a factor
of $\sim 100$).  The resulting sample 
can thus constrain the luminosity function reasonably.
Table~\ref{lf6tab} lists the parameters of the various $z\ga 6$ \lya\
surveys that we consider in our analysis.

The LALA survey (Rhoads et al. 2000, 2003) covers the largest area,
1260 square arcminutes, to a depth of $2 \times 10^{-17}
\ergcm2s$ at $z\approx 6.5$.  Followup spectroscopy of three good
candidates showed that one is a galaxy at $z=6.535$ and the
other two are not (Rhoads et al 2004, hereafter R04).  
Kodaira et al.\ (2003; hereafter K03) use Subaru to search an area of 814
square-arcminutes to a depth of $0.9 \times 10^{-17} \ergcm2s$ and
find 73 candidates. They obtained spectra for 9 of their candidates,
among which two are confirmed as \lya\ on the basis of a single
asymmetric emission accompanied by a continuum drop, which would
imply 16 \lya\ sources among 73 candidates. Two further
sources from K03 show asymmetric lines with a continuum too
faint to ascertain a Lyman break.  These are likely to also be \lya,
resulting in a best estimate of 32 \lya\ sources among the K03 candidates.
The Hu et al.\ (2002) narrowband search for
lensed sources finds intrinsically the faintest $z>6$ \lya\ object
after applying the factor of 4.5 lensing magnification in the smallest area,
$0.46 \sqamin$ with limiting line flux
$3\times 10^{-18} \ergcm2s$. Their larger
area non-lensed upper limits are consistent with other surveys.
The Santos et al (2004) spectroscopic survey for lensed \lya\ emitters
provides the deepest \lya\ emitter search at $z\approx 6.5$ also.
However, at $z>5.6$, Santos et al found no sources and thus provide
only upper limits to the luminosity function.
Kurk et al (2004) have reported the detection of one
$z\approx 6.5$ source in a slitless spectroscopic search at the
VLT.  Their volume coverage is about 10\% those of K03
 and of R04, and their sensitivity ranges from $1$
to $2\times 10^{-17} \ergcm2s$ depending on the line wavelength.

Cuby et al.\ (2003) report one source at z=6.17 found as a continuum
drop source using narrow-band data. We neglect this source in the
subsequent analysis because of the dissimilarity in discovery
methods.
Also, Stern et al (2004) have recently discovered a serendipitous
$z\approx 6.5$ \lya\ source in slit spectroscopy for an unrelated
project.  We leave this source also out of the present paper both
because we came to know of it after most of our analysis was complete
and because it is difficult to assess the true total volume in which
serendipitous $z\approx 6.5$ galaxies might have been found and
reported by extragalactic astronomers worldwide.


The density of \lya\ emitters and its effective uncertainty is easily
calculated for the LALA survey, Hu et al 2004, and Kurk et al 2004,
all of which have complete spectroscopic followup of a small number of
candidates.  For the K03 sample, we need to account for spectroscopic
completeness in two ways.  First, K03 did not consider a
source confirmed as \lya\ unless it showed both line asymmetry and a
continuum break. This, in our opinion, is too stringent, given that
more than half of our $z=4.5$ sources remain undetected in the
continuum due to high equivalent width of the \lya\ line (Malhotra \&
Rhoads 2002) and still show the continuum break in coadded spectra
(Dawson et al. 2004). Such a selection also biases against high
equivalent width objects.  Therefore we include all four of the
Kodaira sources discussed above as \lya\ in our luminosity function
calculations.  However our primary conclusions do not change qualitatively
if we treat only two K03 objects as confirmed.

Second, the uncertainty in the K03 density estimate is
dominated by the small size of the spectroscopic followup.
Applying the same method we used for the $z\approx 5.7$ LALA
and Hu et al samples, we find a variance of $20/9$ and fractional
uncertainty of $37\%$ for the confirmed
sample size.  Adding in the small Poisson uncertainty in the
candidate counts gives a fractional
uncertainty of $39\%$ in the true \lya\ source density.
The equivalent number of objects for a pure Poisson sample would be
$N_{equiv} = 6.57$.  Thus, we treat K03 as having discovered
6.57 sources in an effective volume $20\% = 6.57/(73*4/9)$ of their true 
survey volume for purposes of our fitting code.

\begin{deluxetable}{lllll}
\tablecolumns{5}
\tablewidth{0pc}
\tablecaption{Surveys at $z\approx 6.5$}
\tablehead{
\colhead{Survey ID} & \colhead{Sensitivity}  & \colhead{Volume} & \colhead{Number} &\colhead{$\sigma$} \\
                    & ($\ergcm2s$)           &  (Mpc$^{-3}$)  &   &  }
\startdata
LALA (Rhoads et al. 2004)	      &	$2\times 10^{-17}$   & $2.11 \times 10^5$ &  1   & 1\\
Kurk et al. 2004  & $1.5\times10^{-17}$  & $1.8  \times 10^4$ &  1  & 1\\
Kodaira et al 2003 &	$9\times 10^{-18}$   & $2.2\times 10^5$   & 32.4 & 12.6 \\
Hu et al. 2002    &	$3\times 10^{-18}$   & 110    		  & 1    & 1 \\
Santos et al. 2004 & $2.5\times 10^{-16}$ & $2.7 \times 10^4$  & 0 & \\
	      & $8  \times 10^{-17}$ & 18929              & 0 &\\
	      & $2.5\times 10^{-17}$ & 11943              & 0 &\\
	      & $8\times 10^{-18} $  &  3000              & 0 &\\
	      & $2.5\times 10^{-18}$ &   598              & 0 &\\
	      & $8\times 10^{-19} $  &   78.9             & 0 &\\
	      & $2.5\times 10^{-19}$ &    11.4            & 0 &\\
\enddata
\tablecomments{Surveys used for determining the $z\approx 6.5$
\lya\ luminosity function.
\label{lf6tab}}
\end{deluxetable}

\begin{figure} 
\epsscale{0.8} 
\plotone{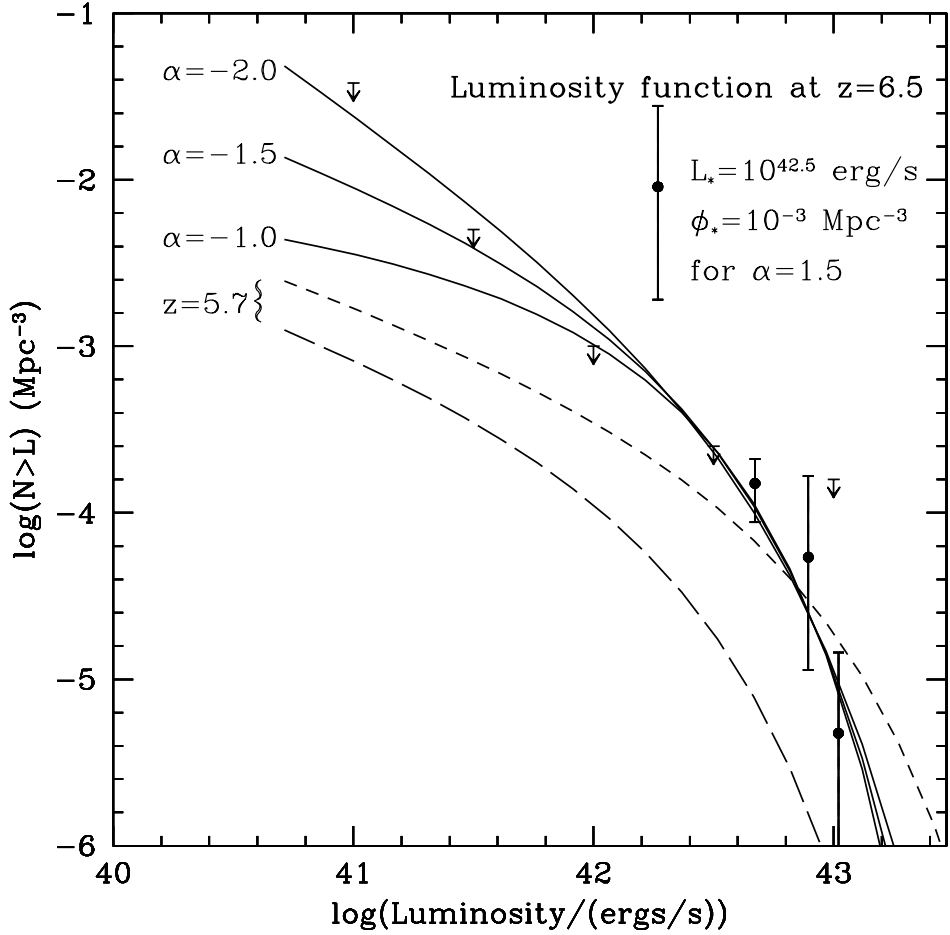} 
\caption{
\lya\ luminosity function at redshift $z\approx 6.5$.  Data points with
error bars show the measured densities above detection threshold and
their corresponding uncertainties for samples from Table~\ref{lf6tab}.  Three
curves are shown, corresponding to the best Schechter function fits to
the data for assumed faint-end slopes $\alpha = 1$, 1.5, and~2 (from
bottom to top). The $z=5.7$ best fit luminosity function is also
consistent with the data (short dashed curve), but $z=5.7$ luminosity
function with flux attenuation of a factor of three (long dashed
curve) is ruled out (see details in section~\ref{rei_sec}).
\label{z6lumfn}}
\end{figure}

We then minimize the Cash parameter for the $z\approx 6.5$ data.
These results are  given in table~\ref{lffittab}, and plotted in
figure~\ref{z6lumfn}. The Hu et al. (2003) lensed object HCM~6A, is
incosistent with Santos et al. upper limits. If the surface density
suggested by the HCM~6a discovery were the norm, Santos et al. (2004)
should have detected 7 \lya\ emitters at $5.6 \la z \la 6.7$.  They
found none.  The corresponding Poisson likelihood is $9\times
10^{-4}$.  Taking now our best fit curve, the chance that Hu et al
would have found HCM~6a is $6.5\%$.  Likely explanations for this
inconsistency include (a) luck, or (b) systematic difficulties in
estimating sensitivity thresholds and volumes, which depend critically
on the details of cluster lens modelling for both Hu et al and Santos
et al.  If the true lensing amplification for HCM~6a were $\sim 1$
(and not $\sim 4.5$ as Hu et al suggest), its intrinsic flux would be
$\sim 2 \times 10^{-17} \ergcm2s$.  The volume covered by Hu et al's
survey to that depth is roughly 400 times the volume stated for an
amplification $>4.5$.  This would move HCM~6A to a brightness and
number density comparable to the LALA and Kurk et al 2004 points.

\begin{deluxetable}{llll}
\tablecolumns{4}
\tablewidth{0pc}
\tablecaption{Fitted luminosity function parameters}
\tablehead{
\colhead{Redshift} & \colhead{$\alpha$}  & \colhead{log$(\lstar/\ergsec)$}
 &  \colhead{log${(\phistar / \Mpc^{-3})}$ }   }
\startdata
5.7 & -1   & -3.6  & 43.45  \\
    & -1.5 & -4.0  & 43.0  \\
    & -2   & -4.8  & 42.75   \\
6.5 & -1   &  -2.9 & 42.5   \\
    & -1.5 &  -3.0 & 42.6  \\
    & -2   & -3.3  & 42.7  \\
\enddata
\tablecomments{Best fit \lya\ luminosity function parameters,
as a function of redshift and faint-end luminosity function slope $\alpha$.
\label{lffittab}}
\end{deluxetable}


\section{Testing for Reionization}
\label{rei_sec}
To test whether \lya\ emission lines from $z\approx 6.5$ galaxies 
are attenuated by the scattering effects of a neutral intergalactic
medium, we compare the observations at  $z\approx 6.5$ to
two hypothetical \lya\ luminosity functions.
In hypothesis (A), we use the best-fitting $z\approx 5.7$ \lya\
luminosity function directly.  This corresponds to the null hypothesis
that the $z\approx 6.5$ IGM neutral fraction, $x_n(z\approx 6.5)$, is
not substantially greater than $x_n(z\approx 5.7)$, and that there has
been no strong evolution of the \lya\ luminosity function.  (Given
that the intervening time is only $\approx 160 \Myr$, or about 17\%
the age of the universe, evolution is likely to be relatively
modest.). These results are tabulated in Table~\ref{rei_table}.  

With the $z=5.7$ luminosity function, LALA should have found 3.6
sources, but found one; Kurk et al. should have found 0.5 sources and
found one, K03 survey should find about 14 sources and the estimate
there is 16--32, depending on whether 2 or 4 of their published
spectra are believed to be \lya\ emitters. We see that the expected
numbers in all surveys are within a factor of 2--3 of the observed
values. Poisson noise aside, we know that variation in number density
of factors of 2--3 are often seen and are due to large scale structure.
 Therefore we conclude that the z=6.5 discoveries are consistent
with the z=5.7 luminosity function.

In model (B), we assume that the \lya\ fluxes are reduced by a factor
of $\sqrt(10)$, which corresponds roughly to the {\it minimum \/}
reduction expected in a neutral IGM for any model considered by Santos
(2004).  For many plausible models, the reduction would be
considerably larger, especially for low-luminosity sources which
cannot carve a significant ionized bubble in the surrounding
IGM. Again the results are tabulated in Table~\ref{rei_table}. With this
luminosity function, LALA and Kurk et al. had 10\% and 3\% chances of
finding a source in their survey, but lucky coincidences do
happen. 

Here, the K03 sample is the most constraining due to its large
size. Under model (B) they should have only 1--2 real \lya\
emitters. They have published spectroscopic followup of 9 out of 73
candidates and 2 to 4 are already specctroscopically confirmed as
\lya\ emitters. We calculate the likelihood of obtaining the observed
spectroscopic confirmations by modelling the number of \lya\ emitters
in the field with a Poisson random variable, and the number of
spectroscopic confirmations achieved with a binomial probability
distribution. This gives a likelihood of $6\%$ for $\ge 4$
spectroscopic confirmations under hypothesis (A) (and $40\%$ for $\ge
2$ confirmations).  Adopting instead hypothesis (B) reduces the
likelihood to $8\times 10^{-5}$ for $\ge 4$ confirmations (and $1.1\%$
for $\ge 2$ confirmations).  Thus imposing a $3\times$ reduction in
\lya\ line fluxes reduces the likelihood for the K03 data by a factor
of $\ga 40$. The Hu et al. galaxy HCM~6A is unlikely to have been
discovered, with 2\% and 0.5\% probability under the two luminosity
function models, and does not fit with a best-fit luminosity function
for z=6.  See section~\ref{lf6sec} for further discussion.

\begin{deluxetable}{lllll}
\tablecolumns{5}
\tablewidth{0pc}
\tablecaption{Comparison of $z\approx 6.5$ data to model luminosity functions}
\tablehead{
\colhead{} & \colhead{LALA} & \colhead{Kurk} & \colhead{Kodaira} & \colhead{Hu}
}
\startdata
Observed &        1  &    1 & 32.4 &    1 \\
$z=5.7$ (raw) & 3.6  & 0.55 & 14.3 & 0.02 \\
$z=5.7$ (0.5 dex attenuated) & 0.10 & 0.029 & 1.7 & 0.005 \\
\enddata
\tablecomments{Comparison between observed and modelled \lya\ galaxy
sample sizes.  The observed number is the candidate count corrected for
spectroscopically determined reliability where appropriate.  The ``z=5.7
(raw)'' line shows the expected number of \lya\ emitters in each survey
assuming that the best fitting $z\approx 5.7$ LF holds for $z\approx 6.5$.
The ``attenuated'' line corresponds to the $z\approx 5.7$ LF with $L_{\star}$
reduced by a factor of $\sqrt{10}$ (and $\phi_{\star}$ unchanged). 
This approximates the effect of a neutral IGM on the observed \lya\ lines
under conservative assumptions.\label{rei_table}}
\end{deluxetable}

\section{Star Formation and Metal Production}
\label{sfr_sec}

We now estimate the star formation rate density in \lya\ galaxies at
$z\approx 5.7$ and $z\approx 6.5$ by integrating over a Schechter
luminosity function. The luminosity density in \lya\ photons is
$L_{\star} \phi_{\star} (\alpha+1)!$ (e.g., Peebles 1993). At
$z\approx 5.7$ and $\alpha=-1.5$, we obtain a luminosity density of
$\ldlya(5.7) = 1.8 \times 10^{39} \ergsec \Mpc^{-3}$, corresponding to
an SFRD of $1.8 \times 10^{-3} \Msun \year^{-1} \Mpc^{-3}$.  At
$z\approx 6.5$ and $\alpha=-1.5$, the best fitting luminosity function
implies a luminosity density of $\ldlya(6.5) = 7\times 10^{39} \ergsec
\Mpc^{-3}$, corresponding to an SFRD of $7 \times 10^{-3} \Msun
\year^{-1} \Mpc^{-3}$.  As discussed in section~\ref{rei_sec}, the
$z\approx 6.5$ data are consistent with the $z\approx 5.7$ LF, which
means that the differences in luminosity and SFR densities are within
the measurement uncertainties.  We therefore adopt $\sim 3 \times
10^{-3} \Msun \year^{-1} \Mpc^{-3}$ as a weighted average estimate for
\lya\ galaxies at $z\sim 6$.  Uncertainties in the luminosity
functions lead to a factor of $\sim 3$ uncertainty in this SFRD.

%

This result substantially exceeds the lower bounds obtained from
the input surveys without integrating over a luminosity function.
For example, K03 find $5.2\times 10^{-4} \Msun \year^{-1} \Mpc^{-3}$, about
1/5 of our value.  A lower bound from the Lyman break galaxies is $\ge 4.7
\times 10^{-4} \Msun \year^{-1} \Mpc^{-3}$ at $5.6<z<6.1$ (Stanway,
Bunker, \& McMahon 2003), consistent with the K03 lower bound and
again below the \lya\ luminosity function integral.  The omission of
galaxies with active star formation but no prominent \lya\ line
(e.g. Malhotra et al. 2004) from the samples we consider can only
increase the $z\approx 6.5$ SFRD above our estimate. Of course, the
stellar population model used to derive the conversion between
\lya\ photons and star formation activity is sensitive to unknown
details of the IMF and metallicity evolution at high redshift, and may
be incorrect.  Comparison with other methods of estimating the SFRD
may be sensitive to these assumptions.

The metal production rate from the \lya\ galaxies at $z\sim 6$
can be estimated because both
\lya\ photons and heavy elements are produced predominantly by the
most massive stars.  We use the relation $d M_Z / d t = 500 L_i /
c^2$, where $L_i \ga 2 \llya$ is the luminosity in ionizing photons
and $M_Z$ is the mass in elements with atomic number $Z \ge 6$ (Madau
\& Shull 1996).  This gives a metal production rate of
of $53 \Msun/\Myr/\Mpc^3 \times {\ldlya}/
(3\times 10^{39} \ergsec \Mpc^{-3})$.  For comparison, the UV
luminosity density in continuum-selected objects at $z\sim 4$
(uncorrected for dust absorption) implies a metal production of $85
\Msun/\Myr/\Mpc^3$ (Madau \& Shull 1996).

Multiplying our metal production rate by the $0.85 \Gyr$ age
of the universe at $z\approx 6.5$ gives an estimate of
$4.5\times 10^4 \Msun/\Mpc^3$ for the mean volume density of metals,
or a mean metallicity estimate of $7 \times 10^{-6}$ (for
$\Omega_{b,0} = 0.048$).  This is about $7\%$ of the metal production
required to reionize the universe with UV photons
produced by stellar nucleosynthesis (Gnedin \& Ostriker 1996). 


\section{Discussion and Conclusions}
\label{finale}

We show that the \lya\ galaxy population at $z\approx 6.5$ shows no
evidence for a neutral intergalactic medium, by comparing the
$z\approx 6.5$ \lya\ survey results to a \lya\ luminosity function
derived at $z\approx 5.7$.  The $z\approx 6.5$ data are fitted by the
$z\approx 5.7$ luminosity function with an acceptable likelihood, but
the alternative hypothesis that all \lya\ fluxes are attenuated by a
factor of $\sim 3$ is much less likely. All but one data points, taken
in different parts of the sky, are consistent with the luminosity
function at that redshift bin, therefore we do not see any evidence of
patchy reionization, which would attenuate \lya\ in one patch and
not the other. Within the limitation of small number of fields considered
here, we do not see any evidence of "reionization in progress". More data
would, of course, help.

We also derive the best fitting Schechter function parameters for
both the $z\approx 5.7$ and $z\approx 6.5$ samples, and use the
results to estimate the star formation rate density ($\sim 3\times 10^{-3}
\Msun \year^{-1} \Mpc^{-3}$) and metal production ($\sim 50 \Msun \Myr^{-1}
\Mpc^{-3}$) in \lya\ galaxies at $z\sim 6$.  These rates imply
that \lya\ emitters contribute $\sim 7 \times 10^{-6}$ to the
mean metallicity of the universe by redshift $6.5$, and
contribute $\sim 7\%$ of the ultraviolet photon budget needed to 
reionize the universe.

The \lya\ test it is sensitive to neutral fractions $x_n$
in the range $0.1 \la x_n \la 1$.  It therefore complements the
Gunn-Peterson test, which ``saturates'' for very small IGM neutral
fractions ($x_n \sim 10^{-4}$ in a uniform IGM, and $x_n \la 10^{-2}$
in any case; e.g., Fan et al 2002).  Our present result implies
$x_n \ll 1$ (and probably $x_n \la 0.3)$ at $z\approx 6.5$.  This
represents the strongest {\it upper} limit presently available
on the neutral fraction at this redshift.  The WMAP result
(Spergel et al 2003) provides an integral constraint on the ionized
column density along the line of sight, but does not place so
strong a constraint on any particular redshift.  By combining all
of these constraints, and other tests in future, it will be possible
to build a much more complete observational history of reionization.

\acknowledgements 
We thank Zoltan Haiman for suggesting we use Cash statistics, and Daniel
Stern for a swift Kit-B which led to a quicker completion of the paper.

\end{document}